\documentclass[conference,letter]{IEEEtran}
\usepackage[utf8]{inputenc}
\usepackage[english]{babel}
\usepackage{amsmath}
\usepackage{amssymb}
\usepackage{bbm}
\usepackage{array}
\usepackage{cite}
\usepackage{cmap}
\usepackage{graphicx}
\usepackage{url}
\usepackage{hyperref}
\usepackage{siunitx}
\usepackage{tikz}
\usepackage[figurename=Fig.]{caption}
\usetikzlibrary{arrows}
\usetikzlibrary{shapes.multipart}

\graphicspath{{pics/}}

\DeclareSIUnit{\dBm}{dBm}

\newcommand\copyrighttext{%
\footnotesize \textcopyright \enspace 2017 IEEE. Personal use of this material is permitted. Permission from IEEE must be obtained for all other uses, in any current or future media, includin
g reprinting/republishing this material for advertising or promotional purposes, creating new collective works, for resale or redistribution to servers or lists, or reuse of any copyrighted component of this work in other works.  DOI: \href{https://doi.org/10.1109/WoWMoM.2017.7974300}{10.1109/WoWMoM.2017.7974300}
}
\newcommand\copyrightnotice{%
\begin{tikzpicture}[remember picture,overlay]
\node[anchor=south] at (current page.south) {\fbox{\parbox{\dimexpr\textwidth-\fboxsep-\fboxrule\relax}{\copyrighttext}}};
\end{tikzpicture}%
}

\begin{document}
\title{Mathematical Model of LoRaWAN Channel Access \thanks{The reported study was partially supported by RFBR, research project No. 15-37-70004 mol\_a\_mos.}}

\author{\IEEEauthorblockN{Dmitry Bankov, Evgeny Khorov and Andrey Lyakhov}
        \IEEEauthorblockN{Institute for Information Transmission Problems, Russian Academy of Sciences, Moscow, Russia\\Email: \{bankov, khorov, lyakhov\}@iitp.ru}
}

\maketitle
\copyrightnotice

\begin{abstract}
While 3GPP has been developing NB-IoT, the market of Low Power Wide Area Networks has been mastered by cheap and simple Sigfox and LoRa/LoRaWAN technologies. Being positioned as having an open standard, LoRaWAN has attracted also much interest from the research community. Specifically, many papers address the efficiency of its PHY layer. However MAC is still underinvestigated. 
Existing studies of LoRaWAN do not take into account the acknowledgement and retransmission policy, which may lead to incorrect results. In this paper, we carefully take into account the peculiarities of LoRaWAN transmission retries and show that it is 
the weakest issue of this technology, which significantly increases failure probability for retries. The main contribution of the paper is a mathematical model which accurately estimates how packet error rate depends on the offered load. In contrast to other papers, which evaluate LoRaWAN capacity just as the maximal throughput, our model can be used to find the maximal load, which allows reliable packet delivery. 

\end{abstract}


\section{Introduction}
LoRaWAN is a relatively new protocol designed to provide cheap and reliable wireless connectivity in various  Internet of Things scenarios.
Being a Low Power Wide Area Network technology operating in the ISM band, it rapidly got popularity in both industry and academic communities. 
Literature review shows that in spite of numerous studies of its PHY layer \cite{centenaro2016long, vangelista2015long, goursaud2015dedicated}, the MAC layer got little attention, even though it has multiple issues \cite{bankov2016limits, mikhaylov2016analysis} that limit its performance.
However, as LoRaWAN is designed to support networks of thousands of devices, it is crucial not only to consider the performance of this technology in point-to-point scenarios, but also to evaluate its applicability in case of highly-populated networks.

To calculate throughput of LoRaWAN networks, in existing studies of the MAC layer (e.g., see \cite{adelantado2017understanding}), the authors typically use the classical approach for modeling ALOHA networks \cite{aloha}. The papers (e.g. \cite{augustin2016study}) also limit the study to unacknowledged mode, which has no control acknowledgements (ACKs). Thus, with no control traffic the throughput increases. However the reliability of transmission decreases.   

In this paper, we provide a mathematical model for a LoRaWAN network operating in the acknowledged mode. We explain why the usage of classical ALOHA-like approach underestimates the collision probability and develop an accurate mathematical model which takes into account LoRaWAN peculiarities related to retransmission policy. 
  

\section{LoRaWAN Channel Access Description}
A typical LoRaWAN \cite{lorawan} network consists of end devices, called \emph{motes}, gateways (GWs), and a server.
Motes are connected to the GWs via wireless LoRa links.
Gateways gather information from the motes, send it to the server via an IP network, and forward packets from the server to the motes.

LoRaWAN devices operate in different ways. Depending on operation, the standard describes three classes of devices. The basic functionality for sporadic uplink data transmission is described as class A operation and is studied in this paper. 

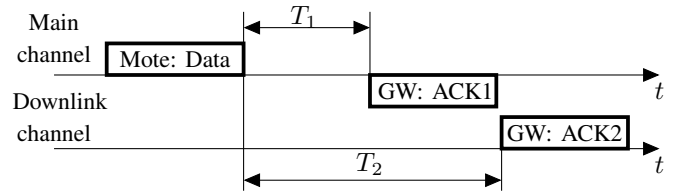
\begin{figure}[!tb]
\centering
\begin{tikzpicture}[scale=0.7]
\draw [arrows={-triangle 45}] (0,2.2) -- (11.5,2.2);
\node [text width=1.1 cm, align=center] at (0,  2.9) {\small{Main channel}};
\node [text width=1.1 cm, align=center] at (0,  1.4) {\small{Downlink channel}};
\draw [arrows={-triangle 45}] (0,0.8) -- (11.5,0.8);
\node at (11.5,  1.9) {$t$};
\node at (11.5,  0.5) {$t$};
\draw [line width=0.5mm] (1, 2.2) rectangle (3.6, 2.8);
\node [text width=1.5cm, align=center] at (2.3,  2.5) {\small{Mote: Data}};
\draw [line width=0.5mm] (6, 1.6) rectangle (8.4, 2.2);
\node [text width=1.6cm, align=center] at (7.25,  1.9) {\small{GW: ACK1}};
\draw [line width=0.5mm] (8.5, 0.8) rectangle (10.9, 1.4);
\node [text width=1.6cm, align=center] at (9.7,  1.1) {\small{GW: ACK2}};
\draw (6,    2) -- (6,   3.4);
\draw (3.6,  0) -- (3.6, 3.4);
\draw (8.5,  0) -- (8.5, 1);
\draw [arrows={triangle 45-triangle 45}] (3.6,3.1) -- (6,3.1);
\draw [arrows={triangle 45-triangle 45}] (3.6,0.2) -- (8.5,0.2);
\node at (4.75,  3.3) {$T_1$};
\node at (6,  0.5) {$T_2$};
\end{tikzpicture}
\caption{LoRaWAN channel access}
\label{fig:channel_access}
\vspace{-1em}
\end{figure}

A LoRaWAN network simultaneously works in several wireless channels. For example, in Europe they can use three main channels and one downlink channel.
To transmit a data frame, each mote randomly selects one of the main channels (see Fig. \ref{fig:channel_access}).
Having received the frame, the GW sends two ACKs. The first one is sent in the main channel, where the frame was received, $T_1$ after frame reception. The second ACK is sent in the downlink channel after timeout $T_2 = T_1 + \SI{1}{\s}$.
If a mote receives no ACK, it makes a retransmission.
The standard recommends making a retransmission in a random time drawn from $[1, 1 + W]$ seconds, where $W = 2$. Note that the recommended $W$ is too small and, as we show in the paper leads to the ``avalanche effect''. 


At the PHY layer, LoRaWAN uses Chirp Spread Spectrum modulation.
Its main feature is that signals with different spreading factors can be distinguished and received simultaneously, even if they are transmitted in the same time on the same channel.
Spreading factor, together with the channel width and the coding rate, determines the data rate.
Lower data rates extend transmission range and improve transmission reliability.
For the first transmission attempt, the rate is determined by the GW.
The standard also recommends decrementing data rate every two consequent transmission failures, limiting the number of retransmissions by $RL = 7$.
The first ACK is sent at a data rate that is lower than the data rate for the frame transmission by a configurable offset (it can be zero).
The second ACK should always be sent at a fixed data rate, by default the lowest one.

\section{Problem Statement} 
\label{sec:scenario}

Consider a LoRaWAN network that consists of a GW and $N$ motes and operates in $F$ main channels and one downlink channel.
The motes use data rates $0, 1, ..., R$, set by the GW. 
Let $p_i$ be the probability that a mote uses data rate $i$.

We consider that a frame collision occurs when two frames are transmitted in the same channel at the same data rate, and they intersect in time.

The motes generate frames according to a Poisson process with total intensity $\lambda$ (the network load).
All motes transmit frames with 51-byte Frame Payload which corresponds to the biggest payload that can fit a frame at the lowest data rate.
The frames are transmitted in the acknowledged mode, and ACKs carry no frame payload.
We consider a situation, when motes have no queue, i.e. if two messages are generated, a mote transmits the most recent one.

For the described scenario, it is important not only to know the nominal channel capacity, but also to find the maximal load at which the network can provide reliable communications. In other words, we need \emph{to find the packet error rate (PER) as a function of network load $\lambda$}.

\section{Mathematical Model}

To solve the problem, we develop a mathematical model of the transmission process.
As the first transmission attempts are described by the Poisson process, to find the PER in these assumptions, in Section \ref{first}, we consider the approach used to evaluate ALOHA networks\cite{aloha} and extended to take into account ACKs.
This approach is however inapplicable for retransmissions, because they do not form a Poisson process, so in Section \ref{retries} we propose another way to take them into account and thus to improve the accuracy of the model.

\subsection{The First Transmission Attempt}
\label{first}
The first transmission attempt is successful with probability
\begin{equation}
\label{eq:success1}
P_{S,1} = \sum_{i = 0}^{R} p_{i} P^{Data}_i P^{Ack}_{i},
\end{equation}
where $P^{Data}_i$ is the probability that the data frame is transmitted without collision at data rate $i$ and $P^{Ack}_i$ is the probability that at least one ACK out of two is received by the mote, provided that the data frame is successful.

Since the packets transmitted in different channels and at different rates do not collide, we need to consider separately each combination of channel and data rate. Specifically for rate $i$ and one of $F$ channels, the load equals $r_i = \frac{\lambda p_i}{F}$.

A data frame transmission is successful if it intersects with no transmission of another frame or an ACK sent by the GW as a response to previous frame.
Let $T^{Data}_{i}$ and $T^{Ack}_i$ be the durations of a data frame and an ACK, respectively, at rate $i$.
Intersection with a frame does not occur if no frames are generated in the interval $[-T^{Data}_{i}, T^{Data}_{i}]$, relative to the beginning of the considered frame.
For a Poisson process of frame generation, such an event happens with probability $e^{-2 r_i T^{Data}_{i}}$. 
We consider that the GW cancels ACK transmission if it is receiving a data frame, so a collision can happen only if the ACK is generated in the interval $[-T^{Ack}_{i}, 0]$.
The rate of ACK generation is $P^{Data}_i r_i$, so the probability to avoid collision with an ACK is $e^{-r_i P^{Data}_i T^{Ack}_{i}}$.
Finally, $P^{Data}_i$ can be found from the following equation:
\[ P^{Data}_i = e^{-(2 T^{Data}_{i} + P^{Data}_i T^{Ack}_{i}) r_i}. \]

As for ACKs, the probability that at least one ACK arrives is calculated according to the inclusion-exclusion principle:
\[ P^{Ack}_i = P^{Ack1}_{i} + P^{Ack2}_i - P^{Ack1}_i P^{Ack2}_i, \]
where $P^{Ack1}_{i}$ and $P^{Ack2}_i$ are the probabilities that the first and the second ACK, respectively, is transmitted successfully, provided that data was transmitted at rate $i$.
The first ACK is transmitted successfully if no data frame intersects it:
\[ P^{Ack1}_i = e^{-\left(\min\left(T_1, T^{Data}_{i}\right) + T^{Ack}_{i}\right) r_i}. \] 
Here we take the minimum of $T^{Data}_{i}$ and $T_1$, because if a frame exceeds $T_1$, it breaks the acknowledged frame, but such an event is already taken into account by $P^{Data}_i$.
The second ACK is transmitted successfully if no data frame is successful in any other channel or at any other data rate, such that its second ACK would intersect the considered one: 
\[ P^{Ack2}_i = e^{-T^{Ack}_{0} \lambda \left(1 - \frac{p_i}{F}\right) \sum_{j = 0}^{R} P^{Data}_j p_j}.\]


\subsection{Retransmissions}
\label{retries}

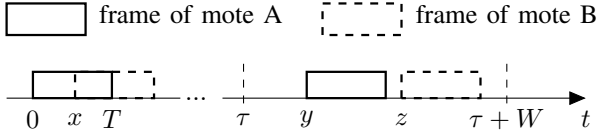
\begin{figure}[!t]
	\centering
	\begin{tikzpicture}[scale=0.7]
	\draw (0,1) -- (3.3,1);
	\draw [arrows={-triangle 45}] (4.1,1) -- (11,1);
	\node at ( 11,  0.6) {$t$};
	\node at (0.5,  0.6) {$0$};
	\node at (1.3,  0.6) {$x$};
	\node at (2.0,  0.6) {$T$};
	\node at (3.6,    1) {$...$};
	\node at (4.5,  0.6) {$\tau$};
	\node at (5.7,  0.6) {$y$};
	\node at (7.5,  0.6) {$z$};
	\node at (9.5,  0.6) {$\tau + W$};
	\draw [line width=0.3mm] (0.5, 1) rectangle (2.0, 1.5);
	\draw [dashed, line width=0.3mm] (1.3, 1) rectangle (2.8, 1.5);
	\draw [line width=0.3mm] (0, 2.2) rectangle (1.5, 2.8);
	\node at (3.5,  2.6) {frame of mote A};
	\draw [dashed, line width=0.3mm] (6.0, 2.2) rectangle (7.5, 2.8);
	\node at (9.5,  2.6) {frame of mote B};
	\draw [dashed] (4.5,  0.9) -- (4.5, 1.8);
	\draw [line width=0.3mm] (5.7, 1) rectangle (7.2, 1.5);
	\draw [dashed, line width=0.3mm] (7.5, 1) rectangle (9.0, 1.5);
	\draw [dashed] (9.5,  0.9) -- (9.5, 1.8);
	\end{tikzpicture}
	\caption{Retransmission}
	\label{fig:retransmission}
\vspace{-0.5em}
\end{figure}

Consider a case, when two motes transmit frames with collision, as shown in Fig. \ref{fig:retransmission}.
Let 0 be the time when the frame of mote A begins, and $x$ be the offset for frame of mote B.
Motes choose a channel for retransmission randomly.  
If they choose different channels, the collision is resolved.
Otherwise, with probability $\frac{1}{F}$, they choose the same channel.
In this case, let $y$ and $z$ be the times when motes A and B start their retransmission, respectively.
The value of $y$ is distributed uniformly in the interval $[\tau, \tau + W]$, where $\tau$ is the frame duration $T$ plus the timeout for the ACK.
The value of $z$ is distributed uniformly in the interval $[\tau + x, \tau + x + W]$.
The retransmission results in a new collision, if $[z, z + T]$ intersects with $[y, y + T]$, which happens with the probability
\begin{align*}
P_x &= \frac{\int\limits_{0}^{T} r_i e^{-r_i x} \int\limits_{0}^{W} \int\limits_{x}^{W + x} \frac{\mathbbm{1}\left(y \leq z \leq y + T\right) + \mathbbm{1}\left(z \leq y \leq z + T\right)}{W^2} dz dy dx}{\int\limits_{0}^{T} r_i e^{-r_i x} dx} =\\
&=\frac{T}{W^2} \left(2 W - \frac{3}{2} T - \frac{2}{T r_{i}^2} + \frac{1}{r_i \tanh(\frac{r_i T}{2})}\right),
\end{align*}
where $\mathbbm{1}(condition)$ is the indicator function which equals 1 if $condition$ is true and 0 otherwise.

Motes have the same probability of being the first and the second one, so the probability that there is no collision equals
\[ P^{Data}_{i, Re} = 1 - 2 P_x / F. \]
The average probability of a successful transmission $P_{S}$ is
\[ P_{S} = P_{1} P_{S, 1} + (1 - P_{1}) P_{S, Re}, \]
where $P_{S, Re}$ is the probability of a successful retransmission, calculated as in eq. \eqref{eq:success1}, using $P^{Data}_{i, Re}$ instead of $P^{Data}_i$, and $P_{1}$ is the probability that the transmission is the first one (not a retry).
$P_{1}$ is reverse to the average number of transmission attempts per a frame:
\[ P_1 = \left(1 + \left(1 - P_{S, 1}\right) \sum\limits_{r = 0}^{RL} \left(1 - P_{S, Re}\right)^r P^{r + 1}_{N}\right)^{-1},\]
where $P_{N} = \sum_{i = 0}^{R} p_i e^{-\frac{\lambda}{N}(T^{Data}_i + T_2 + T^{Ack}_0 + \langle T_{wait} \rangle)}$ is the probability that a new frame does not arrive during the transmission and $\langle T_{wait} \rangle = 1+W/2$ is the average interval that a mote waits before a retransmission.
The packet error rate is calculated as $PER = 1 - P_{S}$.

The model estimates PER correctly up to such network load, that new frames arrive at the motes as quickly as the motes drop the frames due to inability to resolve collisions after $RL$ retransmission attempts.
It means that the load equals
\[ \lambda^* = F \left(\sum_{i = 0}^{R} p_i \left( T^{Data}_{i} + T_2 + T^{Ack}_{0} + \langle T_{wait} \rangle\right) RL \right)^{-1}. \]

\section{Numerical Results}

\begin{figure}[tb]
	\center{\includegraphics[width=0.9\linewidth]{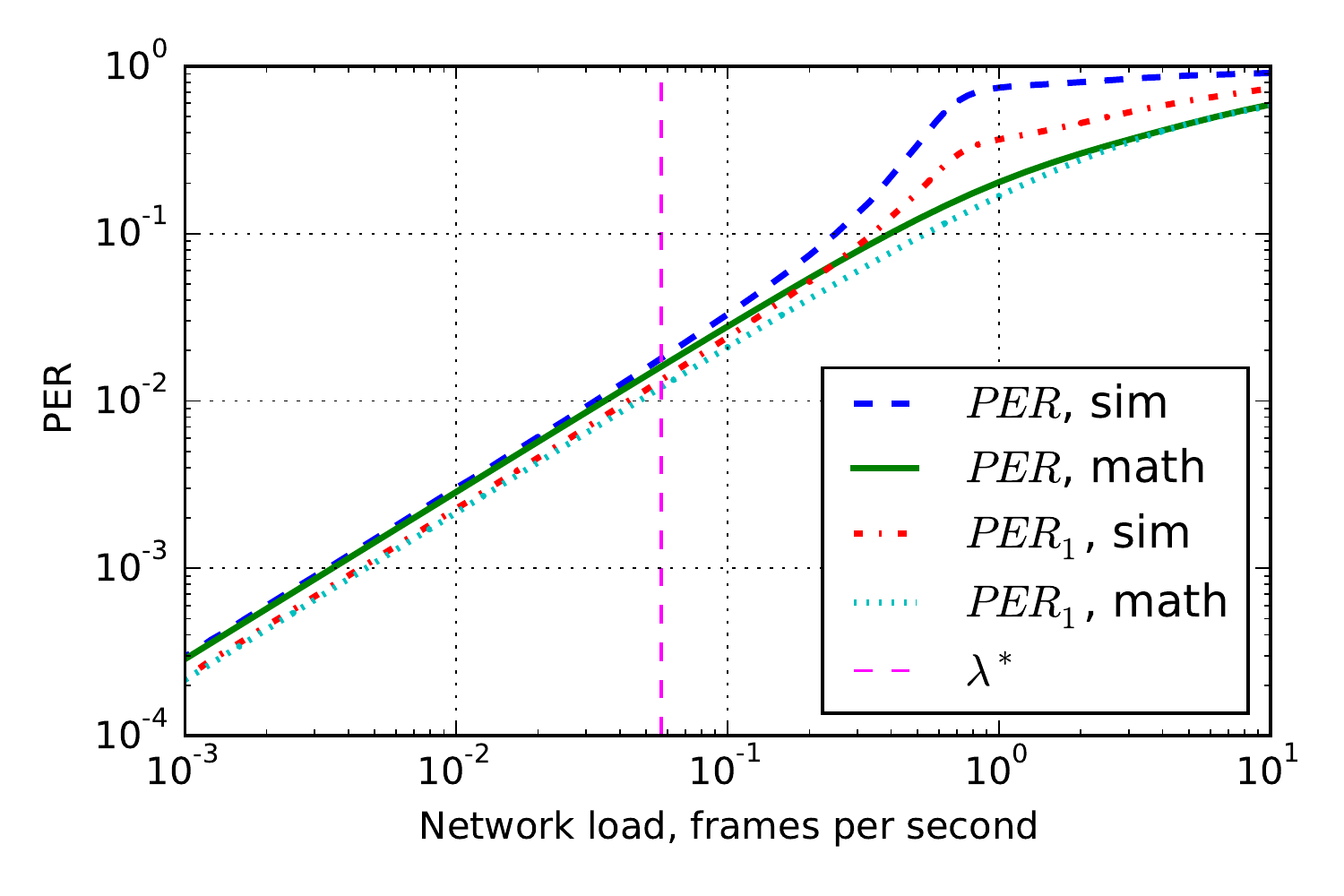}}
	\caption{Dependency of PER on the network load}
	\label{fig:per}
\vspace{-0.5em}
\end{figure}

Let us use the developed model to evaluate performance of a LoRaWAN network. As in \cite{adelantado2017understanding}, we consider a scenario, when the motes are distributed uniformly in a circular area with radius of $\SI{1}{\km}$ around the GW, and the path-loss is described by Okumura-Hata model for urban environment.
We consider EU 863-880 MHz ISM band.
In this case, the data rates are distributed as follows: $p_0 = 0.28, p_1 = 0.2, p_2 = 0.14, p_3 = 0.1, p_4 = 0.08, p_5 = 0.2$.
We simulate a network with 1000 motes and compare the average $\mathrm{PER}$ and  $\mathrm{PER}_1$ for the first transmission attempt with those obtained with the developed mathematical model.
The results are shown in Fig. \ref{fig:per}.
Because of inefficient retransmission parameters the real $\mathrm{PER}$ is by 50\% greater than $\mathrm{PER}_1$. Thus, by taking into account retransmissions, we have significantly improved the accuracy of the model.
From Fig. \ref{fig:per} we also see that we correctly estimate $\lambda^*$ which is the highest load when we can neglect high-order collisions and the ``avalanche effect'' inherent to the default retransmission parameters. 
Non-adaptive and small retransmission window does not allow to resolve collisions with high number of packets, and involving new motes in collisions is faster than packet dropping or collision resolution. This significantly limits the capacity of a LoRaWAN network. While the network can transmit several packets per second, because of a poor retransmission policy the PER rapidly tends to 1, when the load exceeds $10^{-1}$ packets per second.

\section{Conclusion}
\label{sec:conclusion}

In the paper, we develop the first accurate mathematical model of acknowledged uplink transmissions in LoRaWAN networks with class A devices.
We have shown that leaving out of consideration retransmission process significantly overestimates efficiency of a LoRaWAN network.
In contrast, our model takes into account peculiarities of the retransmission process and correctly estimates packet error rate when the load is lower than some threshold $\lambda^*$, which is found in the paper.
However the area with the higher loads is not interesting from a practical point of view.
Indeed, after the load exceeds the described threshold, PER rapidly grows to 1 because retransmissions form an ``avalanche''.
Thus in this area LoRaWAN cannot provide reliable communications. 



\bibliographystyle{ieeetr}
\bibliography{biblio.bib}

\end{document}